\documentstyle[aps,12pt,epsfig,multicol]{revtex}

\begin{document}
\draft
\tighten

\newcommand{\onecolm}{\end{multicols}\vspace{-1.5\baselineskip}
\vspace{-\parskip}\noindent\rule{0.5\textwidth}{0.1ex}\rule{0.1ex}{2ex}
\hfill}
\newcommand{\twocolm}{\hfill\raisebox{-1.9ex}{\rule{0.1ex}{2ex}}
\rule{0.5\textwidth}{0.1ex}\vspace{-1.5\baselineskip} \vspace{-\parskip}
\begin{multicols}{2}}

\title{Analytic stochastic treatment of a nonlinear quantum model 
with negative diffusion}
\author{Roberta Zambrini$^1$ and Stephen M. Barnett$^2$}
\address{$^1$ Instituto Mediterr\'aneo de Estudios Avanzados, 
IMEDEA (CSIC-UIB),\\ Campus Universitat Illes Balears, E-07071 
Palma de Mallorca, Spain. \\ http://www.imedea.uib.es/PhysDept/}
\address{$^2$Department of Physics and Applied Physics, University of Strathclyde,\\ 
107 Rottenrow, Glasgow G4 0NG, Scotland}
\maketitle

\centerline{\today}

\begin{abstract}

We apply a proposal of Yuen and Tombesi, for treating stochastic
problems with negative diffusion, to the analytically soluble
problem of the single-mode anharmonic oscillator.  We find that
the associated stochastic realizations include divergent
trajectories.  It is possible, however, to solve the stochastic
problem exactly, but the averaging must be performed with
great care. 
\end{abstract}

\begin{multicols} {2}


\section{Introduction}
\label{introduction}

The treatment of even quite simple quantum optical systems can
present a significant technical challenge.  
The description of any isolated system can be given using a density 
operator,
with time evolution  governed by the Liouville equation \cite{huang}.
When the system of interest is not isolated, but can exchange
both energy and fluctuations with a surrounding environment,
the evolution of the system density operator is governed by a
master equation \cite{Steve}.
In particular master equations provide a practical method to treat 
such systems but direct solution of these is not usually possible. 
It is often
possible, particularly for problems involving optical field modes,
to map the operator master equation onto a partial differential
equation for a quasi-probability distribution.  It may be possible
to solve this equation or to map it onto an equivalent stochastic
process that can be simulated numerically.

Mapping the quantum problem onto a stochastic system relies on a
formal similarity between the partial differential equation,
obtained from the master equation, and the Fokker-Planck equation
associated with Brownian motion.  The Fokker-Planck equation for
the dynamics of a single field mode or harmonic oscillator is
typically of the form
\begin{eqnarray}
\label{FP} {\frac{\partial}{\partial t}}{\mathcal W} &&=-
{\frac{\partial}{\partial \alpha}} A_\alpha {\mathcal W} -
{\frac{\partial}{\partial \alpha^*}} A_{\alpha^*} {\mathcal W} +\\
&&{\frac{1}{2}}{\frac{\partial^2}{\partial \alpha^2}} D_{\alpha
\alpha}{\mathcal W} + {\frac{1}{2}}{\frac{\partial^2}{\partial
\alpha^{*2}}} D_{\alpha^* \alpha^*}{\mathcal W} +
{\frac{\partial^2}{\partial \alpha
\partial \alpha^*}} D_{\alpha \alpha^*}{\mathcal W},\nonumber
\end{eqnarray}
where ${\mathcal W}$ is the quasi-probability distribution for the
phase-space associated with the mode and parameterized by the
complex variables $\alpha$ and $\alpha^*$ \cite{Steve,GardinerZ}.
The requirement that ${\mathcal W}$ be a real-valued function
imposes the conditions that $A^*_{\alpha^*}=A_\alpha$,
$D_{\alpha^* \alpha^*}=D^*_{\alpha \alpha}$ and $D_{\alpha
\alpha^*}$ is real. This equation can be mapped onto a pair of
stochastic differential equations for the phase-space coordinates
(also written as $\alpha$ and $\alpha^*$) in the form
\begin{eqnarray}
\label{alphastoch}  \dot\alpha & = &  v_\alpha  + \xi (t) \\
\label{alpha*stoch} \dot\alpha^* & = &  v_{\alpha^*} + \xi^* (t),
\end{eqnarray}
where $v_\alpha $ and $v_{\alpha^*}$ are functions of the drift
and diffusion coefficients ($A_i$ and $D_{ij}$) appearing in Eq.
(\ref{FP}) and the dot denotes a derivative with respect to time.
The terms $\xi (t)$ and $\xi^* (t)$ are stochastic fluctuating
terms with correlation functions related to the
diffusion coefficients. There is no unique stochastic
representation of a given Fokker-Planck equation.  In this paper
we work with the Stratonovich form of the stochastic integral
\cite{Gardiner}.  A brief discussion of this is given in Appendix
\ref{SDEs}.

Unfortunately, not all problems of interest can be converted into
the Fokker-Planck form (\ref{FP}).  Systems of interest in quantum
nonlinear optics often produce equations for the evolution of
quasi-probabilities that have derivatives of higher than second
order and it is not known how to treat these within the stochastic
formalism.  The usual approach is to simply drop these terms to
produce ``stochastic electrodynamics".  It has been shown,
however, that this frequently used approximation does not
correctly reproduce higher-order correlations such as those
predicted to occur in parametric oscillators
\cite{DrummondK,Kinsler}. A second, more subtle, problem is that
even when we do obtain an equation of the form (\ref{FP}), it
might still not be possible to map this onto SDEs of the form
(\ref{alphastoch}) and (\ref{alpha*stoch}). The difficulty arises
when we have negative diffusion, that is when $D_{\alpha
\alpha^*}<|D_{\alpha \alpha}|$. With negative diffusion, we are
led to SDEs in which $\xi^*$ cannot be the complex conjugate of
$\xi$ and hence $\alpha^*$ will not be the complex conjugate of
$\alpha$.  It was to resolve problems of this kind that the
positive $P$ representation was introduced
\cite{GardinerZ,Drummond80,Drummond81,Gilchrist}.

In this paper we consider a proposal by Yuen and Tombesi to
convert the evolution equation for the $Q$ quasi-probability into
a pair of SDEs \cite{Yuen,Tombesi}.  Their idea is that the
correct averages should be obtained by {\it formal} application of
the Langevin method by simply ignoring the presence of negative
diffusion.  These authors applied their method to a single-mode
evolving under the influence of a quadratic Hamiltonian in the
presence of damping and showed that this gave the known evolution
for this problem.  In this paper we apply the Yuen-Tombesi
approach to the more demanding, but still analytically soluble
problem of the undamped anharmonic oscillator
\cite{Milburn,Yurke,Tanas}. This model is known to cause
difficulties with stochastic simulations derived from the positive
$P$ representation \cite{GardinerZ,Gilchrist,Plimak}.  We find
that the Yuen-Tombesi method gives the correct results but that it
cannot reliably be applied to numerical simulation of the problem.
We trace the origin of this difficulty to the order in which
stochastic averages and averages over the initial phase-space
distribution have to be performed.

\section{Method of Yuen and Tombesi}
\label{YuenTombesi}

The method of Yuen and Tombesi was designed to deal with problems
in which the evolution equation for the $Q$ function displays
negative diffusion.  The $Q$ function for a single field mode or
oscillator can be written in a number of forms, the simplest of
which is \cite{Steve,GardinerZ,foot}
\begin{eqnarray}
\label{QDefn}Q(\alpha, \alpha^*) = \frac{1}{\pi}\langle \alpha
|\hat\rho| \alpha \rangle,
\end{eqnarray}
where $\hat\rho$ is the density operator for the mode.  This
distribution can be used to obtain anti-normal ordered moments of
the annihilation and creation operators by integration over the
complex $\alpha$ plane:
\begin{eqnarray}
\label{QAntinormal}\langle \hat a^n \hat a^{\dagger m} \rangle =
\int d^2\alpha \ Q\alpha^n\alpha^{*m}.
\end{eqnarray}

We consider systems (such as the anharmonic oscillator) in which
the evolution equation for the $Q$ function is of the form given
in Eq. (\ref{FP}), with {\it negative} diffusion.  This leads to
associated SDEs in which the stochastic variable $\alpha^*(t)$ is
not the complex conjugate of $\alpha(t)$.  As an example, consider
an equation in which $D_{\alpha \alpha^*}=0$.  This necessarily
implies negative diffusion associated with $D_{\alpha \alpha}$ and
$D_{\alpha^* \alpha^*}$.  We can follow the method outlined in
Appendix \ref{SDEs} to obtain a pair of SDEs that are equivalent
to our evolution equation for $Q$ \cite{foot2}:
\begin{eqnarray}
\label{alphastochex} \dot\alpha & = & A_\alpha -
\frac{1}{4}\frac{\partial}{\partial \alpha}D_{\alpha \alpha} +
\sqrt{D_{\alpha \alpha}}f_\alpha  \\ \dot\alpha^* & = & A_\alpha^*
- \frac{1}{4}\frac{\partial}{\partial \alpha^*}D_{\alpha^*
\alpha^*} + \sqrt{D_{\alpha^* \alpha^*}}f_{\alpha^*}.
\label{alpha*stochex}
\end{eqnarray}
It might appear that these equations are mutual complex conjugates
but this is not the case as the two Gaussian noise terms are {\it
independent} and hence do not take complex conjugate values.  It
follows that we cannot interpret $\alpha$ and $\alpha^*$ as mutual
complex conjugates.  The situation is reminiscent of that found
with the positive $P$ representation and we employ the same
notation by writing our stochastic variables as $\alpha(t)$ and
$\alpha^+(t)$ \cite{Drummond80}.  Anti-normal ordered expectation
values should then correspond to stochastic averages of
corresponding functions of $\alpha$ and $\alpha^+$, with $\hat
a(t)$ replaced by $\alpha(t)$ and $\hat a^\dagger(t)$ replaced by
$\alpha^+(t)$.

We can introduce the variables $\alpha$ and $\alpha^+$ more
formally by means of the complex function
\begin{eqnarray}
\label{Q+} \tilde{Q}(\alpha, \alpha^+)=\frac{1}{\pi}\langle
0|e^{\alpha^+\hat a}\hat\rho e^{\alpha \hat a^\dagger}|0\rangle
e^{-\alpha^+\alpha},
\end{eqnarray}
which is a function of $\alpha$ and $\alpha^+$ but not of their
complex conjugates.  This reduces to the familiar $Q$ function
(\ref{QDefn}) when $\alpha^+ = \alpha^*$.  We can convert our
master equation for $\hat\rho$ into an evolution equation for
$\tilde{Q}$ by making the substitutions
\begin{eqnarray}
\hat \rho \hat a^\dagger & \rightarrow & \left(\alpha^+ +
\frac{\partial}{\partial \alpha} \right)\tilde{Q} \nonumber \\
\hat a^\dagger \hat \rho & \rightarrow & \alpha^+ \tilde{Q} \nonumber
\\\hat  a \hat \rho & \rightarrow & \left(\alpha +
\frac{\partial}{\partial \alpha^+} \right)\tilde{Q} \nonumber
\\ \hat \rho \hat a & \rightarrow & \alpha \tilde{Q}. \label{Q+subs}
\end{eqnarray}
The resulting equation for $\tilde{Q}$ will be of the same form as
that for our $Q$ function with $\alpha^*$ replaced by $\alpha^+$.

\section{Anharmonic Oscillator}
\label{Anharmonic}

The anharmonic oscillator is one of the simplest analytically
solvable models in quantum optics.  The Hamiltonian for this model
can be written in the form
\begin{eqnarray}
\label{anharmHam} \hat H = \omega\left(\hat a^\dagger\hat a +
\frac{1}{2} \right) + \mu(\hat a^ \dagger\hat a)^2,
\end{eqnarray}
where $\omega$ is the natural angular frequency for the mode and
we work with units in which $\hbar=1$.  The term proportional to
$\mu$ is sometimes written in normal order as $\mu \hat a^\dagger
\hat a^\dagger \hat a \hat a$.  This is the same model but with
the $\omega$ changed to $\omega + \mu$.  It is convenient to
remove the free evolution of the mode and this can be achieved by
working in an interaction picture rotating at angular frequency
$\omega$.  The interaction picture form of the Hamiltonian
(\ref{anharmHam}) is
\begin{eqnarray}
\label{anharmIntHam}\hat H_I = \mu(\hat a^ \dagger\hat a)^2.
\end{eqnarray}
This Hamiltonian has been used in quantum optics as a model for
the Kerr nonlinear refractive index.  Despite its simplicity, it
produces a number of nonclassical effects including squeezing
\cite{Tanas} and Schr\"odinger cat states \cite{Yurke}, that is
superpositions of coherent states.  The accurate reproduction of
these features, especially the cat states, presents a stiff
challenge to a stochastic simulation method such as that proposed
by Yuen and Tombesi \cite{Yuen,Tombesi}.  The fact that the model
is analytically soluble means that we can compare the results of
such simulations with exact analytical expressions.  We will give
an example of this comparison in the following section.  In this
section we present a brief review of some of the known features of
the model.

It is clear from the form of the Hamiltonian that it commutes with
the number operator $\hat a^ \dagger\hat a$.  It follows that the
number of excitations (or photons) in the mode will be conserved
and that the photon number states $|n\rangle$ will be the
eigenstates of our interaction Hamiltonian
\begin{eqnarray}
\label{eigenstates}\hat H_I |n\rangle = n^2 \mu |n\rangle.
\end{eqnarray}
The corresponding time-evolution operator is
\begin{eqnarray}
\label{U}\hat U(t)=\exp(-i \hat H_I t) =
\sum_{n=0}^\infty|n\rangle\langle n|e^{-in^2\mu t}
\end{eqnarray}
and it follows that the evolution of our oscillator will be
periodic with period $2\pi/\mu$.  If we can expand our initial
state in terms of the number states, then we can use this result
to solve for the time-evolved state in the Schr\"odinger picture.
For example, an initial coherent state $|\alpha_0\rangle$ will
evolve to the state
\begin{eqnarray}
\label{evolveCoh}\hat
U(t)|\alpha_0\rangle=e^{-|\alpha_0|^2}\sum_{n=0}^\infty
\frac{\alpha_0^n}{\sqrt{n!}} e^{-in^2\mu t}|n\rangle.
\end{eqnarray}
This state has a rich structure that can be seen in pictures of
the associated quasi-probability distributions
\cite{Milburn,Tanas}. The state has a simple form at the quarter
periods when it can be written as \cite{Yurke}
\begin{eqnarray}
\hat U[\pi/(2\mu)]|\alpha_0\rangle & = &
\frac{1-i}{2}|\alpha_0\rangle + \frac{1+i}{2}|-\alpha_0\rangle
\nonumber \\ \hat U[\pi/\mu]|\alpha_0\rangle & = &
|-\alpha_0\rangle \nonumber \\ \hat U[3\pi/(2\mu)]|\alpha_0\rangle
& = & \frac{1+i}{2}|\alpha_0\rangle +
\frac{1-i}{2}|-\alpha_0\rangle . \label{quarter}
\end{eqnarray}
The state at one quarter and three quarters of a period is a
superposition of the coherent states $|\alpha_0\rangle$ and
$|-\alpha_0\rangle$. Such superposition states have interesting
nonclassical properties and have been called Schr\"odinger cat
states.

Our stochastic treatment is designed to produce expectation values
of operators for the oscillator.  These can also be calculated
analytically, but this is most easily performed in the Heisenberg
interaction picture.  
The time-evolved annihilation and creation operators are
\begin{eqnarray}
\label{evolveda}
\hat a(t) & = & \hat U^\dagger(t)\hat a  \hat U(t)
=  e^{-i\mu t} e^{-i2\mu t \hat a^\dagger \hat a}\hat a
\\  \label{evolvedadagg} \hat a^\dagger(t) & = & \hat U^\dagger(t)
\hat a^\dagger \hat U(t) =  
e^{i\mu t}\hat a ^\dagger e^{i2\mu t \hat
a^\dagger \hat a},
\end{eqnarray}
where we have written the initial operators as $\hat a$ and $\hat
a^\dagger$.  It is straightforward to use these expressions to
calculate expectation values for functions of $\hat a(t)$ and
$\hat a^\dagger (t)$.  For example, the expectation value of the
annihilation operator for the coherent state $|\alpha_0\rangle$ is
\begin{eqnarray}
\label{aexpect}\langle \hat a (t) \rangle &&= e^{-i\mu t}\langle
\alpha_0 | e^{-i2\mu t \hat a^\dagger \hat a}\hat a
|\alpha_0\rangle = \\
&&e^{-i\mu t}\alpha_0
\exp\left(|\alpha_0|^2(e^{-i2\mu t}-1)\right).\nonumber
\end{eqnarray}
In this expression we have omitted the free-evolution
in the form of a factor $e^{-i\omega t}$.  This corresponds to working
in a frame rotating at frequency $\omega$, associated with our choice of
interaction picture.  
All expressions in this paper will be given in this frame.
The expectation value of $\hat a^\dagger (t)$ is the complex
conjugate of Eq. (\ref{aexpect})  and higher order moments can also be calculated
without difficulty.

The evolution equation for the $Q$ function can be written in the
form \cite{Milburn}
\begin{eqnarray}
\frac{\partial}{\partial t}Q =&& -i\mu
\left[\frac{\partial}{\partial \alpha^*}(2|\alpha|^2-3)\alpha^*Q -
\frac{\partial}{\partial \alpha}(2|\alpha|^2-3)\alpha Q \right.
\nonumber \\ \label{QFP}&&+\left.
\frac{\partial^2}{\partial \alpha^{*2}}\alpha^{*2}Q -
\frac{\partial^2}{\partial \alpha^2}\alpha^2Q  \right].
\end{eqnarray}
Comparison with Eq. (\ref{FP}) reveals that this equation has
negative diffusion ($D_{\alpha \alpha^*}=0<|D_{\alpha
\alpha}|=2|\mu \alpha^2|$) and hence is a good candidate with
which to test the ideas of Yuen and Tombesi.  We should emphasize
that the partial differential Eq. (\ref{QFP}) itself does not 
present any difficulties
in spite of the negative diffusion \cite{V+R}. 
Indeed we can solve this equation directly to give the
correct $Q$ function \cite{Milburn}.

\section{Analytic stochastic treatment of the anharmonic oscillator}
\label{Analytic}

We can re-express the evolution of our $Q$ function, given by Eq.
(\ref{QFP}) as an equivalent stochastic process using the method
outlined in Appendix \ref{SDEs}.  A simple and natural choice is
to set $C_{\alpha \alpha^*}=0=C_{\alpha^* \alpha}$ so that
$C_{\alpha \alpha}=\sqrt{i2\mu}\alpha$ and $C_{\alpha^*
\alpha^*}=\sqrt{-i2\mu}\alpha^*$.  The evolution equation for our
$Q$ function clearly displays negative diffusion and so we write
our SDEs in terms of the variables $\alpha$ and $\alpha^+$.  For
the choices described above, our SDEs become
\begin{eqnarray}
\label{SDEalpha}
\dot{\alpha}& = &-i\mu 2(\alpha^+\alpha-1)\alpha + \xi \alpha  \\
\label{SDEalpha+} \dot{\alpha}^+& = &i\mu
2(\alpha^+\alpha-1)\alpha^+ + \xi^+ \alpha^+,
\end{eqnarray}
where $\xi$ and $\xi^+$ are complex, white Gaussian noises with
the stochastic averages
\begin{eqnarray}
\nonumber
\langle \xi(t)\xi(t')\rangle_S & = & 2i\mu\delta(t-t') \\
\nonumber
\langle \xi^+(t)\xi^+(t')\rangle_S & = & -2i\mu\delta(t-t') \\
\langle \xi^+(t)\xi(t')\rangle_S & = & 0.
\label{xicorrns}
\end{eqnarray}
We will require averages over both the stochastic noise 
realizations and also
over the initial quasi-probability distribution.  The subscript
$S$ identifies the fact that we have carried out the stochastic
average.
The stochastic averages (22) do not fully determine the forms
of the noise terms.  It is clear, however, that $\xi^+(t)$ cannot
be the complex conjugate of $\xi (t)$.  It has been suggested
that the considerable freedom in choosing the forms of $\xi (t)$
and $\xi^+(t)$ can be used to suppress, although not completely
remove, stochastic sampling errors in the analogous problem in
the positive $P$ representation.  The analysis presented in this
section is independent of this choice of Gaussian noise and
hence the freedom to select the forms of $\xi (t)$ and $\xi^+(t)$
will not address the problem uncovered.

We require the solution of Eqs. (\ref{SDEalpha}) and
(\ref{SDEalpha+}) with the initial conditions $\alpha(0)=\beta$
and $\alpha^+(0)=\beta^*$.
These mean that
$\alpha^+(0) = \alpha^*(0)$ and allow us to use the initial $Q_0$ function
to perform the average over the initial state.
 As already noted in Sect.
\ref{YuenTombesi}, the form of the stochastic noise means that
$\alpha^+(t)$ will not take the value $(\alpha^*(t))$ in any given
realization.  The full quantum average will only be obtained by
performing an average over the $Q_0$ function for the initial state.
For the coherent state $|\alpha_0\rangle$ this is
\begin{eqnarray}
\label{initQ} Q_0(\beta)=\frac{1}{\pi}e^{-|\beta-\alpha_0|^2}.
\end{eqnarray}
We denote the average obtained by integrating over $\beta$ by the
subscript $Q_0$:
\begin{eqnarray}
\nonumber\langle F\left(\alpha(t),\alpha^+(t)\right)\rangle_{Q_0} =
\int_{-\infty}^{\infty}d^2\beta
Q(\beta)F\left(\alpha(t),\alpha^+(t)\right).
\end{eqnarray}
Quantum expectation values should be obtained on performing both
the stochastic average and the average over the this $Q_0$ function.
In particular, the mean value of $\hat a$ at time $t$ will be
\begin{eqnarray}
\langle \hat{a}(t) \rangle = \langle \langle \alpha (t) \rangle_S
\rangle_{Q_0}. \label{GenMeana}
\end{eqnarray}
We have not yet given a prescription for the order, if any, in
which these averages must be performed.  We will see that this
question is of some importance for the solution of the SDEs.

In this section we will calculate the expectation value of the
annihilation operator at time $t$ by solving the SDEs
(\ref{SDEalpha}) and (\ref{SDEalpha+}).  We start by noticing that
the combination $\alpha^+\alpha$ satisfies the equation
\begin{eqnarray}
\label{alpha+alpha}\frac{d}{dt}\alpha^+\alpha = (\xi +
\xi^+)\alpha^+\alpha.
\end{eqnarray}
The formal solution to this equation is
\begin{eqnarray}
\label{alpha+alphaSoln}\alpha^+(t)\alpha(t) = \beta^*\beta
e^{\int_0^t[\xi(t')+\xi^+(t')]dt'}.
\end{eqnarray}
This already suggests that the stochastic simulation of this
problem may run into difficulties.  We expect the average obtained
from $\alpha^+(t)\alpha(t)$ will be $\langle\hat a(t)\hat
a^\dagger(t)\rangle$, which should take the constant value
$|\alpha_0|^2+1$.  The solution (\ref{alpha+alphaSoln}), however,
clearly shows that the stochastic noise will cause
$\alpha^+(t)\alpha(t)$ to fluctuate away from its initial value
in a single realization of the stochastic process.
The average is  constant but the corresponding
variance increases in time.
The presence of a complex driving force $\xi+\xi^+$
means that $\alpha^+(t)\alpha(t)$ can acquire any complex value.
Nevertheless
we can proceed by inserting our solution (\ref{alpha+alphaSoln})
into our SDEs (\ref{SDEalpha}) and (\ref{SDEalpha+}).  We find
that the resulting equations are linear. In particular, the
equation for $\alpha(t)$ becomes
\begin{eqnarray}\nonumber
\dot{\alpha}(t) = \left[-i2\mu
\left(|\beta|^2e^{\int_0^t[\xi(t')+\xi^+(t')]dt'}-1
\right)+\xi(t)\right]\alpha(t),
\end{eqnarray}
the solution of which is
\onecolm
\begin{eqnarray}
\label{alphaSol}\alpha(t)= \beta\exp\left\{\int_0^t\left[-i2\mu
\left(|\beta|^2 e^{\int_0^{t'} [\xi(t'')+\xi^+(t'')]dt''}-1
\right)+\xi(t')\right] dt'\right\}.
\end{eqnarray}
Similar expressions have been given for the same model treated in
the positive $P$ representation \cite{Plimak}.  The average of
this quantity should be the expectation value of $\hat{a}(t)$. Let
us start by performing the stochastic average. This can be
achieved most readily by expanding the outer exponential in powers
of $|\beta|^2$
\begin{eqnarray}
\langle \alpha(t) \rangle_S = \beta e^{i2\mu t} \langle
e^{\int_0^t \xi(t')dt'}\left[1 - i2\mu|\beta|^2\int_0^t
e^{\int_0^{t'} [\xi(t'')+\xi^+(t'')]dt''}dt' - \right. \nonumber \\ \left. 
2\mu^2|\beta|^4\int_0^tdt'\int_0^tdt''e^{\int_0^{t'}
[\xi(s)+\xi^+(s)]ds} e^{\int_0^{t''} [\xi(s')+\xi^+(s')]ds'} +
\cdots \right]\rangle_S. \label{alphaStochAv}
\end{eqnarray}
Here we have made explicit use of the Gaussian nature of our
stochastic noise in evaluating the averages of exponential
functions of the noise.  We can evaluate the average of each term
in turn. The order zero and order one terms are
\begin{eqnarray}
\label{firstAv} \langle e^{\int_0^t \xi(t')dt'}\rangle_S & = &
e^{i\mu
t} \\
-i2\mu|\beta|^2\langle
\int_0^tdt'e^{\int_0^{t'}[2\xi(s)+\xi^+(s)]ds}e^{\int_{t'}^t\xi(s')ds'}\rangle_S
& = & -i2\mu|\beta|^2\int_0^tdt'e^{4i\mu t'}e^{-i\mu
t'}e^{i\mu(t-t')}\nonumber \\
& = & -|\beta|^2 e^{i\mu t}(e^{2i\mu t}-1). \label{secondAv}
\end{eqnarray}
\twocolm
It is straightforward to show that the stochastic average of the
term of order $|\beta|^{2n}$ is $(-1)^n|\beta|^{2n}e^{i\mu t}(e^{2i\mu t
}-1)^n/n!$.  It is tempting to re-sum the series in Eq.
(\ref{alphaStochAv}) to give
\begin{eqnarray}
\label{re-sum}\langle \alpha(t) \rangle_S = \beta e^{i3\mu t}\exp
\left(|\beta|^2(1-e^{i2\mu t})\right).
\end{eqnarray}
Let us see the consequences of this re-summing by completing our
calculation of the expectation value of $\hat{a}(t)$ with the
average over $\beta$.  This procedure leads to the expression
\begin{eqnarray}
\nonumber\langle\langle\alpha(t)\rangle_S\rangle_{Q_0} =
\frac{1}{\pi}\int d^2\beta e^{-|\beta-\alpha_0|^2}\beta e^{i3\mu
t}\exp \left(|\beta|^2(1-e^{i2\mu t})\right).
\end{eqnarray}
Inspection of the integrand reveals a problem.  It is clear that
the integrand is {\it unbounded} (and the integral {\it
undefined}) for times $t$ such that $\cos(2\mu t) \leq 0$. 
It is interesting to note that this includes the times,
$\pi /(2\mu)$ and $3\pi /(2\mu)$, at which the anharmonic oscillator
evolves into the Schr\"odinger cat states given in Eqs. (\ref{quarter}).
The problem is that we have assumed that it is acceptable to perform
the stochastic average before performing the average over initial
conditions.  In fact this is not the case and we should perform
the $Q_0$ average first. We can see this by evaluating the average
over the $Q_0$ function before re-summing the series in our
stochastic average given in Eq. (\ref{alphaStochAv}).  This gives
the final average value
\begin{eqnarray}
\langle\langle\alpha(t)&&\rangle_{Q_0}\rangle_S   = \nonumber \\ =&& e^{i3\mu
t}\sum_{n=0}^\infty \frac{(1-e^{i2\mu t})^n}{n!} \int
\frac{d^2\beta}{\pi}\beta^{n+1}\beta^{*n}e^{-|\beta-\alpha_0|^2}
\nonumber \\
 = && e^{i3\mu t}\sum_{n=0}^\infty(1-e^{i2\mu t})^n \sum_{l=0}^n
\frac{(n+1)!}{l!(l+1)!(n-l)!}\alpha_0^{l+1}\alpha_0^{*l} \nonumber \\
 = && e^{i3\mu t}\sum_{l=0}^\infty \frac{|\alpha_0|^{2l}}{(l+1)!}
\sum_{n=l}^\infty (1-e^{i2\mu t})^n \frac{(n+1)!}{(n-l)!l!}
\nonumber \\
 = && e^{-i\mu t}\alpha_0 \exp\left(|\alpha_0|^2(e^{-i2\mu
t}-1)\right),\label{aAvRight}
\end{eqnarray}
which we recognize as the correct answer given in Eq.
(\ref{aexpect}).  Other moments can be obtained in the same manner.

We can see the origin of the incorrect stochastic average given in
Eq. (\ref{re-sum}) by considering the form of the annihilation
operator in the Heisenberg picture, Eq. (\ref{evolveda}), which we
can also write in the form
\begin{eqnarray}
\label{evolvedaanti}\hat a(t)=e^{3i\mu t}\hat a e^{-i2\mu t\hat
a^\dagger \hat a} =e^{3i\mu t}\hat a \vdots e^{(1-e^{i2\mu
t})\hat a \hat a^\dagger}\vdots 
\end{eqnarray}
where $\vdots \ \vdots$ denotes antinormal ordering and we have
used the antinormal ordering theorem for the exponential of $\hat
a^\dagger \hat a$ \cite{Steve}.  We note that this becomes the
expression (\ref{re-sum}) obtained by performing the stochastic
average, if we identify $\hat a$ and $\hat a^\dagger$ with $\beta$
and $\beta^*$ respectively.  We have written Eq.
(\ref{evolvedaanti}) in antinormal order because the $Q$ function
gives antinormally ordered moments.  If we use this expression to
calculate the expectation value of $\hat a(t)$, for our initial
coherent state, then we find
\begin{eqnarray}
\label{antiAv}\langle \hat a(t) \rangle = e^{3i\mu t}\langle
\alpha_0 |\hat a \vdots \exp\left((1-e^{i2\mu t})\hat a \hat
a^\dagger\right)\vdots |\alpha_0 \rangle.
\end{eqnarray}
We can, of course, evaluate this expectation value by putting the
operator into normal ordered form and using the fact that the
coherent states are right-eigenstates of the annihilation
operator.  Our aim, however, is to investigate the problems with
the stochastic average associated with simulations designed to
reproduce antinormal ordered averages.  We can work with the
antinormal ordered form in Eq. (\ref{antiAv}) by expanding the
exponential as a Taylor series and inserting the identity in the
form of an integral over the coherent states $|\beta\rangle$
\cite{Steve}:
\onecolm
\begin{eqnarray}
\label{antiAvInt}\langle \hat a(t) \rangle =e^{3i\mu
t}\langle \alpha_0 | \sum_{n=0}^\infty \frac{(1-e^{i2\mu
t})^n}{n!}\hat a^{n+1} \int
\frac{d^2\beta}{\pi}|\beta\rangle\langle\beta|\hat
a^{\dagger n} |\alpha_0 \rangle 
= e^{3i\mu t}\sum_{n=0}^\infty \frac{(1-e^{i2\mu
t})^n}{n!}\int \frac{d^2\beta}{\pi}\beta
|\beta|^{2n}e^{-|\beta-\alpha_0|^2}.
\end{eqnarray}
\twocolm
Clearly it would be wrong to evaluate the summation before
carrying out the integral.  Evaluating the integral first
corresponds, in our stochastic treatment, to averaging over
initial conditions before performing the stochastic average and
gives the correct result.

It is interesting to note that there is a strong similarity between
the SDEs discussed here and those found for the anharmonic
oscillator in the positive $P$ representation.  Indeed, if we write
equations for $\alpha e^{-i\mu t}$ and $\alpha^+ e^{i\mu t}$,
then we recover the equations discussed by Plimak {\it et al} 
\cite{Plimak}.
An important difference, however, is that the diffusion for the
positive $P$ representation occurs with the opposite sign to that
for the $Q$ function.  This means that the stochastic 
averages (\ref{xicorrns})
have opposite signs when applied to the positive $P$ representation.
We can use the methods described in this section 
to obtain the
expectation value of $\hat a (t)$ in the positive $P$ representation.
The stochastic average gives $\langle \alpha (t) \rangle_S = \beta
e^{-i\mu t} \exp\left( |\beta|^2 (e^{-i2\mu t} - 1)\right)$.
Performing the average of this over a $\delta$-function positive
$P$ distribution, peaked at $\beta = \alpha_0 = \beta^{+*}$,
gives the correct result (\ref{aexpect}).  The positive $P$ representation
is associated with operator moments in normal order and this
seems to be the reason for the well-behaved form of the
stochastic averages for initial coherent states.

\section{Stochastic simulation of the anharmonic oscillator}
\label{Stochastic}

In this section we present results of numerical
 si\-mu\-la\-tions \cite{simulation} of the stochastic process $\alpha(t)$ given in 
Eq. (\ref{alphaSol}).
Our simulations were performed using two
discrete Gaussian processes $\eta_l,\eta_l^+$ of the form
\begin{eqnarray}
\eta_l=\int_{\Delta t}dt'\xi(t')~,~~~~~~\eta_l^+=\int_{\Delta t}dt'\xi^+(t')
\end{eqnarray}
where $t=l\Delta t$.
In this way $<\eta_l\eta_{l'}>=2i\mu{\Delta t}\delta_{ll'}$ and 
$<\eta_l^+\eta_{l'}^+>=-2i\mu{\Delta t}\delta_{ll'}$.
We note that the relations (\ref{xicorrns}) 
do not completely specify the two independent $ complex $ white noises. 
As recently shown in \cite{Plimak} the degree of freedom in the choice of the noise
could be used to improve the results of the numerical simulation 
by choosing the 
stochastic  processes $\xi$ and $\xi^+$ so as to inhibit 
(but not completely suppress)  the fast growth of  
$\alpha(t)$.  In this paper, however, we have considered only the forms
$\xi=\sqrt{2i\mu}\phi$ and  $\xi^{+}=\sqrt{-2i\mu}\phi^{+}$
with $\phi$ and $\phi^{+}$ being $real$ white noises.

Each stochastic realization must start from a single point in phase
space.  For this reason, the analysis of the preceding section leads
us to conclude that diverging trajectories, exploring large values of
$|\alpha|$ are inevitable.  These divergences are responsible for the
unbounded average obtained by performing the stochastic average
before the average over the initial $Q$ distribution.
Each of our simulations starts with at a point $\alpha(0) = 
\alpha^{+*}(0) = \beta$.  
Naturally, the average over the initial $Q$ distribution requires
stochastic realizations for a range of values of $\beta$, weighted
by the distribution (\ref{initQ}).  Consideration of a single value of
$\beta$, however, suffices to illustrate the divergences associated
with individual trajectories.
We observe, in each case, a divergence after some time.  
We can see the origin of these divergences in the SDEs 
(\ref{SDEalpha}) and (\ref{SDEalpha+});
the complex variables $\alpha$ and $\alpha^+$ are not constrained to
be complex conjugate quantities and so, in any given realization, the
combination $\alpha^+\alpha$ can acquire an imaginary part.  This
leads to exponential growth of $\alpha$ or $\alpha^+$.
The time at which this divergence appears varies 
between realizations and also depends on the initial conditions.  In 
particular, the divergence appears earlier for larger values of $|\beta|^2$.
This is because of the exponential dependence of $\alpha(t)$ on $|\beta|^2$
as seen in Eq. (\ref{alphaSol}).

If we select a sufficiently small value of $\beta$ and perform an average 
over a large number of trajectories then we find a result that is, for
short times, in good agreement with the analytical average Eq. (\ref{re-sum}). 
In Fig. \ref{fig0} we have plotted the time evolution of  $Re\langle\alpha(t)\rangle_S$,
obtained by considering $50000$ trajectories, starting
from the initial condition $\beta=0.001+i0.1$ (diamonds line).  For comparison
the analytical expression for the stochastic average is represented by a continuous line. 
At very short times, we observe a near perfect agreement between the numerical results
and the analytical expression.  At longer times, this agreement is lost 
because of the divergence induced by the independent stochastic noises. 

The trajectories start from a single point in phase space.  This 
corresponds to selecting, in each simulation, a $\delta$ function 
phase-space probability distribution.  Such a narrow distribution for the 
$Q_0$ distribution does not correspond to any physically allowed state 
\cite{GardinerZ}.  Indeed, the evolution obtained from the Fokker-Planck 
equation for such an initial condition is highly singular.  It is this 
behavior that is reflected in the divergent numerical simulations.
Fig. \ref{fig1} depicts the numerically obtained value of $\langle\alpha(t)\rangle_S$.
We see that this average explores an extended region of the complex plane.
The analytical average, Eq. (\ref{re-sum}), is represented by the 
small circle.

The relationship between the time at which trajectories diverge and
the initial condition ($\beta$) means that an ensemble of trajectories 
starting from a range of different initial conditions will rapidly produce
divergences.  For this reason the analytical result (\ref{aexpect}) cannot be 
reproduced numerically in any straightforward manner. 

\section{Conclusion}
\label{Conclusion}

In conclusion we have considered a proposal of Yuen and Tombesi \cite{Yuen}
to give a stochastic representation 
of a Fokker-Planck equation with negative diffusion for the $Q$ representation.
We have shown that the
correct analytical moments for an anharmonic oscillator (associated with a
$non$-$linear$ $\chi^{(3)}$ process)
can be obtained from the SDE's.
These results, however, are highly sensitive to the order 
in which averages over the stochastic realizations and
over the distribution of the initial conditions are performed.
It is clear that more sophisticated techniques are required for stochastic
simulation of the problem.  Recent work suggests that the effects of
divergences can be significantly suppressed but not yet eliminated 
\cite{Plimak,Carusotto}.

The system studied in this paper is highly idealized.  We could 
include the effects of loss and expect that these will improve the
stability of the numerical results.  Such an improvement has been
noted in studies of the positive {\it P} \cite{Gilchrist}.
It is possible, however, that there are other interesting systems
that are less sensitive to the noise and for these, stochastic 
simulations using the Yuen-Tombesi method may prove to be a useful
technique.  Possible systems for study in quantum optics include
the Optical Parametric Oscillator and Second Harmonic Generation,
that could be successfully studied with this approach.
Our preliminary studies suggest that there are regimes of operation,
including the threshold, in which the probability for a divergent 
trajectory to occur is very small.  In this case, numerical 
simulation does give stable results.  We will return to this topic
elsewhere.

\acknowledgements

We are grateful to Pere Colet, Emilio Hernandez Garcia, Gian-Luca
Oppo, Maxi San Miguel and Raul Toral for their suggestions and
encouragement. This work was supported through the European Commission
projects QSTRUCT (Project No. ERB FMRX-CT96-0077) and  
QUANTIM (IST-2000-26019).  
SMB thanks the Royal Society of Edinburgh and
the Scottish Executive Education and Lifelong Learning Department
for financial support.

\appendix
\section{}
\label{SDEs}

In this Appendix we present a brief discussion of the link between
a given Fokker-Planck equation and an equivalent stochastic system
(a more complete account can be found in \cite{Gardiner,maxi+raul}). As we
have already noted, the Fokker-Planck equation does not correspond
to a unique stochastic system and so it is natural to start with a
stochastic system. Consider the pair of (Langevin) SDEs
\begin{eqnarray}
\label{alphastochApp}
{\frac{\partial \alpha}{\partial t}} & = & B_\alpha  +
C_{\alpha \alpha}f_{\alpha}(t) + C_{\alpha \alpha^*}f_{\alpha^*}(t)\\
\label{alpha*stochApp} {\frac{\partial \alpha^*}{\partial t}} & =
& B_{\alpha^*} + C_{\alpha^* \alpha^*}f_{\alpha^*}(t) +
C_{\alpha^* \alpha} f_{\alpha}(t),
\end{eqnarray}
with white Gaussian noise terms $f_{i}$, defined to have zero
average and second moments of the form
\begin{eqnarray}
\label{stochavApp} \langle f_{i}(t)f_{j}(t') \rangle =
\delta_{ij}\delta (t-t')
\end{eqnarray}
and the subscripts $i,j$ denote $\alpha$ and $\alpha^*$.

The formal solution of equations (\ref{alphastochApp}) and
(\ref{alpha*stochApp}) is:
\begin{eqnarray}
\nonumber
\alpha(t)&=&\alpha(0)+\int_0^t dt' B_\alpha(t')+\nonumber\\&& \int_0^t
C_{\alpha \alpha}dW(t') + \int_0^t  C_{\alpha \alpha^*}dW_*(t')\\
\alpha^*(t)&=&\alpha^*(0)+\int_0^t dt' B_\alpha^*(t')+ \nonumber\\ &&\int_0^t
C_{\alpha^* \alpha}dW(t') + \int_0^t  C_{\alpha^*
\alpha^*}dW_*(t'), \label{formal.sol}
\end{eqnarray}
where we have introduced the Wiener processes $dW(t)=f_{\alpha}(t)dt$
and $dW_*(t)=f_{\alpha^*}(t)dt$.

In order to use these stochastic processes, we need to give a
prescription for carrying out the stochastic integrals over the
Wiener processes. In this paper, we choose the Stratonovich
interpretation of the stochastic integral in which
\begin{eqnarray}\nonumber
\int_0^t  g(\alpha,\alpha^*)&&dW(t')=\sum_i[W(t_i)-W(t_{i-1})]\cdot
\\ \nonumber &&
g\left(\frac{\alpha(t_i)+\alpha(t_{i-1})}{2},
\frac{\alpha^*(t_i)+\alpha^*(t_{i-1})}{2}\right).
\end{eqnarray}
The reason of this choice, instead of the It\^{o} interpretation,
is that we will be constructing analytical averages over the
stochastic process and the Stratonovich for\-ma\-lism allows us to use
the familiar rules of calculus.

From the Langevin equations it is possible to obtain a {\it
unique} Fokker-Planck equation for the probability distribution
${\mathcal W}(\alpha(t),\alpha^*(t),t)$.  If we consider the
trajectory obtained in a single realization of the stochastic
process $\vec{f}=(f_{\alpha},f_{\alpha^*})$ and start from the
initial value $\vec{\alpha}(0)=(\alpha(0),\alpha^*(0))$, then the
solution at time $t$ is completely determined and the probability
distribution for it is the $\delta$-function
$\delta\left(\alpha-\alpha(t;\vec{\alpha}(0);[\vec{f}(t)])\right)
\delta\left(\alpha^*-\alpha^*(t;\vec{\alpha}(0);[\vec{f}(t)])\right)$.
Considering a set of initial conditions $\vec{\alpha_0} $,
distributed according some initial distribution
$p_0=p(\vec{\alpha}(0);0)$, we can obtain the shape of the
distribution at time $t$:
\begin{eqnarray} \label{pp0}
&& p(\vec{\alpha},t;[\vec{f}(t)] )=\\\nonumber
&&\langle \delta\left(\alpha-\alpha(t;\vec{\alpha}(0);[\vec{f}(t)])\right)
\delta\left(\alpha^*-\alpha^*(t;\vec{\alpha}(0);[\vec{f}(t)])\right)
\rangle_{p_0},
\end{eqnarray}
where the subscript $p_0$ denotes an average over the initial
probability distribution.  The quantity $p$ satisfies the
conservation equation
\begin{eqnarray}
\frac{\partial}{\partial t}p+\frac{\partial}{\partial
\alpha}(\dot\alpha p) +\frac{\partial}{\partial
\alpha^*}(\dot\alpha^*
 p)=0. \label{conserve}
\end{eqnarray}

The complete probability distribution  is obtained by also averaging 
over the stochastic trajectories obtained with different noise
realizations, denoted by the subscript $[\vec{f}(t)]$:
\begin{eqnarray}
{\mathcal W}(\alpha,\alpha^*,t)=\langle
p(\alpha,\alpha^*,t;[\vec{f}(t)] ) \rangle_{[\vec{f}(t)]}.
\end{eqnarray}
The time evolution for the distribution  ${\mathcal W}$ can be
obtained using the continuity equation and gives \cite{Gardiner,maxi}:
\begin{eqnarray}
{\frac{\partial}{\partial t}}{\mathcal W} =
-{\frac{\partial}{\partial \alpha}} B_{\alpha }{\mathcal W} -
{\frac{\partial}{\partial \alpha^*}} B_{\alpha^*}{\mathcal W} +
{\frac{1}{2}}\sum_{ijl}\left[{\partial}_iC_{ij}{\partial}_lC_{lj}\right]{\mathcal
W}, \nonumber
\end{eqnarray}
where the subscripts $i,j,k$ again denote $\alpha$ and $\alpha^*$.
If compare this form of the Fokker-Planck equation with the Eq.
(\ref{FP}) then we obtain the correspondences:
\begin{eqnarray}
\label{identificationsApp}
  &&\label{Dalphaalpha}D_{\alpha \alpha} = C^2_{\alpha \alpha}+ C^2_{\alpha \alpha^*}\\
&&\label{Dalpha*alpha*}D_{\alpha^* \alpha^*}  =  C^2_{\alpha^* \alpha^*}+ C^2_{\alpha^* \alpha}\\
&&\label{Dalphaalpha*}D_{\alpha \alpha^*}  =  C_{\alpha \alpha}C_{\alpha^* \alpha}+C_{\alpha^* \alpha^*}C_{\alpha \alpha^*}\\
 &&A_\alpha =  B_\alpha + {\frac{1}{2}} \left[
\left({\frac{\partial}{\partial \alpha}} C_{\alpha \alpha}\right)C_{\alpha \alpha}+
\left({\frac{\partial}{\partial \alpha^*}} C_{\alpha \alpha}\right)C_{\alpha^* \alpha}+\right.\nonumber\\ 
\label{Aalpha}&&\left.\left({\frac{\partial}{\partial \alpha}} C_{\alpha \alpha^*}\right)C_{\alpha \alpha^*}+
\left({\frac{\partial}{\partial \alpha^*}} C_{\alpha \alpha^*}\right)C_{\alpha^* \alpha^*}
\right]  \\
&&A_{\alpha^*}  =  B_{\alpha^*} + {\frac{1}{2}} \left[
\left({\frac{\partial}{\partial \alpha}} C_{\alpha^*\alpha}\right)C_{\alpha \alpha}+ 
\left({\frac{\partial}{\partial\alpha^*}} C_{\alpha^* \alpha}\right)C_{\alpha^* \alpha}+ \right. \nonumber\\
\label{Aalpha*}&&\left.\left({\frac{\partial}{\partial \alpha}} C_{\alpha^*\alpha^*}\right)C_{\alpha \alpha^*}+
\left({\frac{\partial}{\partial \alpha^*}} C_{\alpha^*\alpha^*}\right)C_{\alpha^* \alpha^*} \right].
\end{eqnarray}
Note that these equations do not give unique forms for the $B$ and
$C$ functions.  This is a consequence of the lack of a unique
stochastic representation for a given Fokker-Planck equation.

If our stochastic variables $\alpha$ and $\alpha^*$ are to be
complex conjugate quantities, then it follows from equations
(\ref{alphastochApp}) and (\ref{alpha*stochApp}) that
$B^*_{\alpha^*} = B_\alpha$, $C^*_{\alpha \alpha}=C_{\alpha^*
\alpha}$ and $C^*_{\alpha \alpha^*}=C_{\alpha^* \alpha^*}$.  These
conditions necessarily imply positive diffusion as, from
(\ref{Dalphaalpha}) to (\ref{Dalphaalpha*}), $D_{\alpha
\alpha^*}>|D_{\alpha \alpha}|$. It follows that the stochastic
variables cannot be complex conjugate quantities when we have
negative diffusion.  In order to avoid possible confusion, we
replace the stochastic variable $\alpha^*(t)$ by $\alpha^+(t)$
whenever there is negative diffusion.

\begin{figure}
\centerline{\epsfysize=6cm \epsfbox{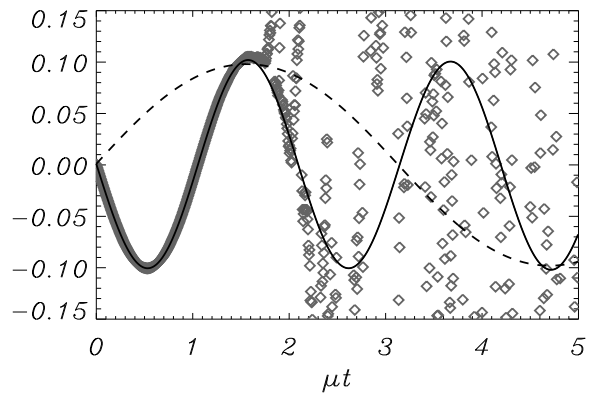}} 
\caption[]{Time e\-vo\-lu\-tion of the sto\-cha\-stic a\-ve\-ra\-ge $Re\langle\alpha(t)\rangle_S$,
using $50000$ trajectories and starting from $\beta = 0.001 + i0.1$.
The diamonds represent numerical values.
The continuous line represents the
analytical result for $\langle\alpha(t)\rangle_S$.
The dashed line represents 
$\langle\langle\alpha(t)\rangle_{Q_0}\rangle_S$
for the initial coherent state $|\alpha_0>$ with
$\alpha_0= 0.001 + i0.1$.} \label{fig0}
\end{figure}
\begin{figure}
\centerline{\epsfysize=7cm \epsfbox{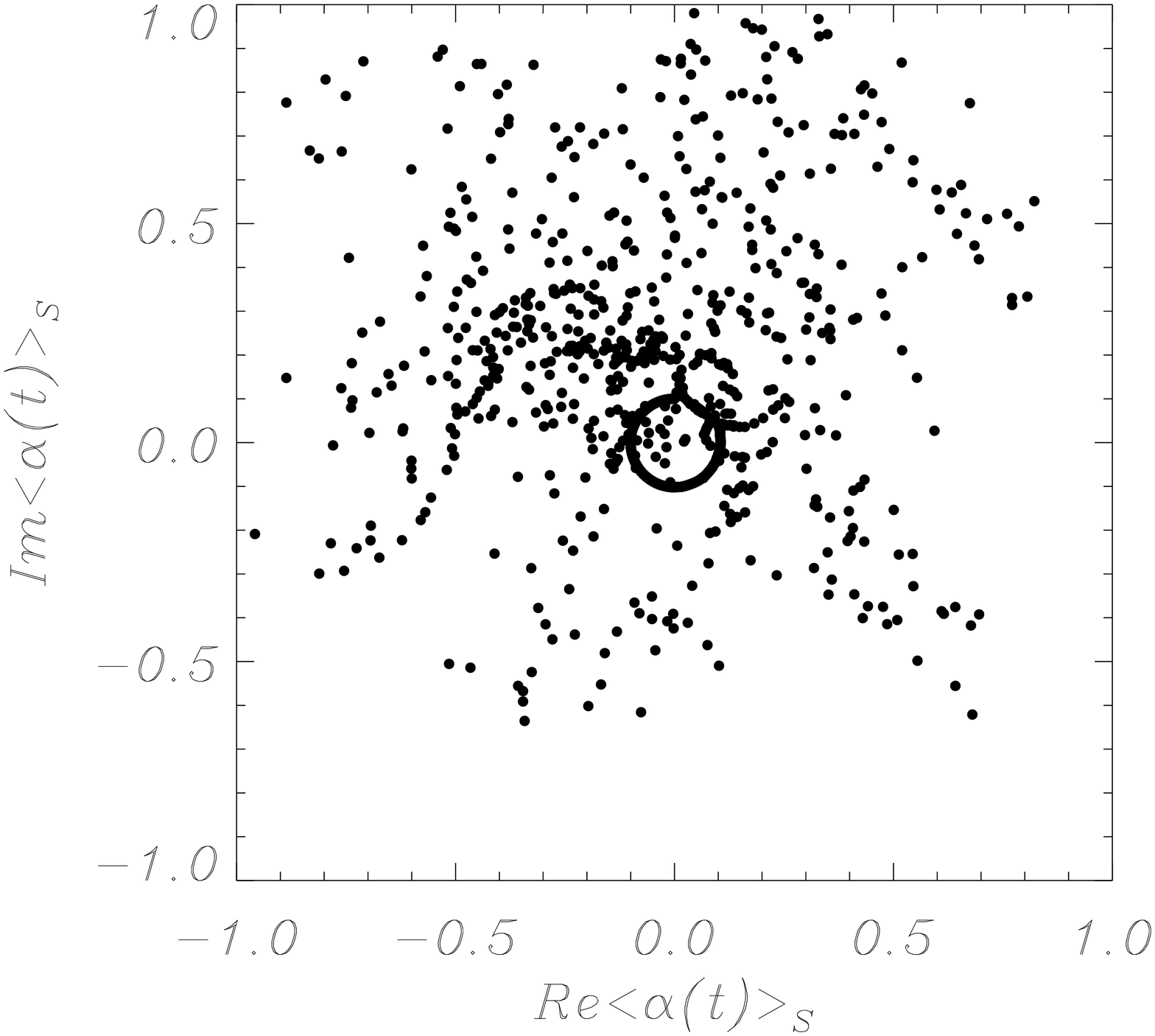}} 
\caption[]{
Phase space
plot of the numerical average $\langle\alpha(t)\rangle_S$, 
the real part of which is
shown in Fig. \ref{fig0}.
The points represent numerical values at different times.
The circle represents the
analytical result for $\langle\alpha(t)\rangle_S$.} \label{fig1}
\end{figure}
\end{multicols}

\begin{thebibliography}{99}

\bibitem{huang} K. Huang, {\it Statistical mechanics}, 2nd ed.
(John Wiley \& Sons, New York, 1987)
\bibitem{Steve} S. M. Barnett and P. Radmore,
{\it Methods in Theoretical Quantum Optics},
 (Oxford University Press, Oxford, England, 1997)
\bibitem{GardinerZ} C. W. Gardiner and P. Zoller, {\it Quantum Noise}
Second Edition (Springer-Verlag, Berlin, 2000)
\bibitem{Gardiner} C. W. Gardiner, {\it Handbook of Stochastic
Processes} (Springer-Verlag, Berlin, 1985)
\bibitem{DrummondK} P. D. Drummond and P. Kinsler, Quant.
Semiclass. Opt. {\bf 7}, 727 (1985)
\bibitem{Kinsler} P. Kinsler, Phys. Rev. A {\bf 53}, 2000 (1996)
\bibitem{Drummond80} P. D. Drummond and C. W. Gardiner, J. Phys.
A: Math. Gen. {\bf 13}, 2353 (1980)
\bibitem{Drummond81} P. D. Drummond, C. W. Gardiner and D. F.
Walls, Phys. Rev. A {\bf 24}, 914 (1981)
\bibitem{Gilchrist} A. Gilchrist, C. W. Gardiner and P. D.
Drummond, Phys. Rev. A {\bf 55}, 3014 (1997)
\bibitem{Yuen} H. P. Yuen and P. Tombesi, Opt. Comm. {\bf 59}, 155 (1986)
\bibitem{Tombesi} P. Tombesi, Phys. Lett. A {\bf 132}, 241 (1988)
\bibitem{Milburn} G. J. Milburn, Phys. Rev. A {\bf 33}, 674 (1986)
\bibitem{Yurke} B. Yurke and D. Stoler, Phys. Rev. Lett. {\bf
57}, 13 (1986)
\bibitem{Tanas} R. Tan\'as, A. Miranowicz and S. Kielich, Phys.
Rev. A {\bf 43}, 4014 (1991)
\bibitem{Plimak} L. I. Plimak, M. K. Olsen and M. J. Collett,
Phys. Rev. A {\bf 64}, 025801 (2001)

\bibitem{foot}This definition differs by a factor of $\pi$ from
that used by Milburn \cite{Milburn}.
\bibitem{foot2} These are obtained from the equations
given in Appendix \ref{SDEs} by the choosing $C_{\alpha
\alpha^*}=0=C_{\alpha^* \alpha}$.

\bibitem{V+R} K. Vogel and H. Risken, Phys. Rev. A {\bf 39}, 4675 (1989)
\bibitem{simulation} Numerical simulation are performed using the Gaussian 
random number generator proposed in: R. Toral, A. Chakrabarti,
Comp. Phys. Comm., {\bf 74}, 327 (1993).

\bibitem{Carusotto} I. Carusotto, Y. Castin, and J. Dalibard,
 Phys. Rev. A {\bf 63}, 023606 (2001).

\bibitem{maxi+raul} M. San Miguel, R. Toral,
``Stochastic Effects in Physical Systems'', in 
{\it Instabilities and Nonequilibrium Structures VI}, 
eds. E. Tirapegui, J. MartÌnez, and R. Tiemann, 
Kluwer Academic Publishers, 35 (2000).
\bibitem{maxi} J. M. Sancho, M. San Miguel, S. L. Katz,  J. D. Gunton,
Phys. Rev. A {\bf26}, 1589 (1982)

\end{thebibliography}
\end{document}